\documentstyle[psfig]{l-aa}

\begin{document}

\thesaurus{3(4.19.1;4.3.1;13.18.2;13.25.3)}

\title {Radio-loud active galaxies in the northern ROSAT All-Sky Survey}
\subtitle {I: Radio identifications}

\author{S.\, A.\, Laurent-Muehleisen\inst{1}
\and R.\, I.\, Kollgaard\inst{1}
\and P.\, J.\, Ryan\inst{1}
\and E.\, D.\, Feigelson\inst{1}
\and W.\, Brinkmann\inst{2}
\and J.\, Siebert\inst{2}}

\institute{Department of Astronomy \& Astrophysics, The Pennsylvania State
University, University Park, PA, USA 16802 \and Max-Planck Institut f\"ur
Extraterrestrische Physik, D-85740 Garching, Germany}

\offprints{salm@astro.psu.edu}

\date{Received date; accepted date}
\maketitle

\begin{abstract}
We present 5\,GHz high resolution VLA observations of 2,127 radio-- and
X-ray--emitting sources found in both the Green Bank (GB) 5\,GHz radio catalog
and the ROSAT All-Sky Survey (RASS).  We report core flux densities and
positions accurate to $\pm$0.5${\arcsec}$.  Combined with the GB measurements
of the total radio emission, we derive the core-to-lobe ratio of objects in
our sample and discuss their core-dominance relative to samples of radio
galaxies and BL Lacertae objects.  Our results show the RASS/Green Bank (RGB)
sample is approximately an order of magnitude more core-dominated than the
radio galaxy sample, but is more than an order of magnitude less
core-dominated than highly beamed BL Lacertae objects.  Using simple beaming
models, this indicates the typical object in the RGB catalog exhibits moderately
beamed radio emission and is oriented at an angle to the line-of-sight
$\overline{\theta}_{\rm RGB}$$\sim$25$^{\circ}$-35$^{\circ}$.  The case of
the origin of the X-ray emission is not as clear; the data are consistent with
either an anisotropic unbeamed or moderately beamed X-ray component.

Tables 2 and 3 which present the RGB catalog are available in their entirety
only from the CDS via anonymous ftp to cdsarc.\linebreak[0]u-strasbg.fr
(130.79.128.5), via the WWW at http://cdsweb.\linebreak[0]u-strasbg.fr/
\linebreak[0]Abstract.html, or at ftp://ftp.\linebreak[0]astro.\linebreak[0]psu.
\linebreak[0]edu\linebreak[0]/pub\linebreak[0]/edf.


\end{abstract}

\section{Introduction} 
The catalog created from the ROSAT All-Sky Survey (RASS) consists of
$\sim$60,000 sources making it the deepest complete sample of soft X-ray
(0.07-2.4 keV) sources ever constructed (\cite{voges:full}).  Previous X-ray
surveys such as the HEAO-1 Large Area Sky Survey (\cite{wood:full}), the
Einstein Extended Medium Sensitivity Survey (\cite{gioia:full};
\cite{stocke:full}), and the Einstein Slew Survey (\cite{elvis:full}),
consisted of less than 5,000 sources total.  Identification programs of these
earlier surveys showed the majority of objects are extragalactic consisting
mainly of quasars, Seyferts, BL Lacertae objects, clusters of galaxies and
occasionally normal galaxies.  Similarly, the bulk properties of various samples
of optically identified objects in the RASS (e.g.\, \cite{molonglo:full};
\cite{Brinkmann:fullids}; \cite{bade:full}) have shown that the RASS contains
thousands of extragalactic objects and will therefore provide the largest
flux-limited sample of X-ray--emitting AGN for the foreseeable future.

Because RASS positions are known to only $\sim$30${\arcsec}$ accuracy, complete
identification of the entire RASS catalog is an enormous task.  Correlations
with deep surveys at other wavelengths can efficiently create subsamples of
manageable size and also select objects of particular interest.  We present
here 2,127 sources which appear in both the RASS and the 1987 Green Bank (GB)
5\,GHz radio survey (\cite{gnc:full}; \cite{gnc2:full}, hereafter called
\cite{gnc2:g96}).  Because $>$70\% of the sources in this RASS-Green Bank
(RGB) sample are optically unidentified and the positional accuracy of both
surveys is low, we obtained high resolution radio observations to enable the
identification of unique optical counterparts.  Out detection of compact core
radio components of these radio-loud active galaxies together with GB
observations of the total radio emission, permits study of the beaming
characteristics of these RASS sources. The multiwavelength properties of the
previously optically identified sources appear in
\cite{Brinkmann:dateids_paperI} and multiband radio observations of a subset are
given by \cite{neumann:date}.

This paper is organized as follows: the construction of the RGB catalog is
discussed in \S2 and the new radio data presented in \S3.  In \S4 we compare
our results to the Green Bank data.  Section 5 uses simple beaming models to
characterize the radio emission of the sources in our catalog and \S6 discusses
the X-ray beaming properties.  The broadband multifrequency properties of the
entire RGB catalog, including newly identified optical counterparts and X-ray
properties, will be presented in \cite{siebert:firstpaperIII}.

\section{The RASS-Green Bank (RGB) Sample}
Analysis of the Green Bank radio data was performed by fitting single elliptical
Gaussian surfaces to local enhancements (\cite{neumann:full}).  The final
catalog contains $\sim$150,000 small diameter sources above a detection
threshold of 3$\sigma$.  The flux density limit is $\sim$15~mJy in the
declination range from $30^{\circ}-75^{\circ}$ and increases to $\sim$24~mJy at
0$^{\circ}$ declination.  In order to identify as many radio-emitting objects in
the RASS as possible, this list was purposely constructed to be deeper than
those lists published later (\cite{becker:full}; \cite{gnc:full}), which are
restricted to $>$5$\sigma$ sources.  The initial radio source catalog is
therefore likely to contain many faint spurious sources.  The rate of spurious
coincidences in the overall RASS-Green Bank correlation ought to be much less,
however, since close proximity (100${\arcsec}$) to a detected X-ray source is
required.  The 1$\sigma$ positional accuracy of the \cite{neumann:date} Green
Bank catalog is approximately $\pm$15${\arcsec}$ for bright sources and
$\pm$40${\arcsec}$ for fainter ones (\cite{neumann:full}).  Direct comparison
of the reanalyzed GB flux densities with those published in \cite{gnc2:g96}
shows that for sources $>$100~mJy the flux densities are accurate to
$\sim$20\% while for sources $<$100~mJy the flux densities differ by up to
$\sim$40\%, with the reanalyzed values generally being higher.

The RASS data were processed using the semi-automatic Standard Analysis
Software System (SASS; \cite{voges92:full}) which determines the sky
coordinates and energy of each photon by applying an aspect solution and
calibration to each X-ray event.  The data are then analyzed using various
source detection algorithms, comprising two sliding window techniques and a
maximum-likelihood method.  Further details are given in \cite{voges:date}.

The RASS and GB surveys were cross-correlated producing a catalog of 2,127
sources with separations of $<$100${\arcsec}$.  These sources are listed in
Tables 2-4 below.  The distribution of the separations is well represented by a
Gaussian with $\sigma$$\sim$17${\arcsec}$ for distances up to 40${\arcsec}$ and
is approximately constant beyond this (Figure 1, B95).  Because of the highly
non-Gaussian nature of the distribution, sources with RASS/GB separations of up
to 100${\arcsec}$ were included in this initial catalog.  The number of spurious
coincidences is believed to be less than 200 objects (B95).  The radio flux
densities range from 15~mJy to 60~Jy and the X-ray fluxes range from
8$\times$10$^{-14}$ to 4$\times$10$^{-10}$~erg~s$^{-1}$~cm$^{-2}$.

\section{New VLA Observations} 
High resolution observations of the RGB sample were made with the NRAO's Very
Large Array\footnote{NRAO is operated by Associated Universities, Inc., under
cooperative agreement with the National Science Foundation.} (VLA) between
October 1992 and September 1995.  The observations were recorded with the
two standard 50~MHz bandwidth IFs at an effective frequency of 4.885~GHz.
Table 1 summarizes the observing parameters including the epoch, array
configuration, average exposure time per source and beam size.  On October 3,
1992, data were collected while the VLA was in a hybrid A/D configuration.  We
were able to obtain flux densities and positions for these sources only by
using the antennas in the low-resolution D-like configuration which yielded
insufficient positional accuracy for unambiguous optical identification.  For
completeness we list these sources separately but do not consider them part of
our well-defined sample.  They are excluded from further analysis.  The region
of the sky covered by the D-configuration observations is approximately
defined by 0$^{\rm hr}$$<$$\alpha$$<$15$^{\rm hr}$,
0$^{\circ}$$<$$\delta$$<$40$^{\circ}$ and 15$^{\rm hr}$$<$$\alpha$$<$16$^{\rm
hr}$, 0$^{\circ}$$<$$\delta$$<$15$^{\circ}$, although high resolution
observations for several objects in this region were obtained.

\begin{table}
\caption[]{Observing Log}
\begin{flushleft}
\begin{tabular}{lrlccclr}
\hline\noalign{\smallskip}
\multicolumn{3}{c}{Date} && Obs & VLA &\multicolumn{1}{c}{Exp.}& Beam \\
& & && Code & Config & (min) & (${\arcsec}$) \\
\noalign{\smallskip}
\hline\noalign{\smallskip}
Oct  &  19  &  1992  &&  a  &   A   & 1.0  & 0.4  \\
May  &   7  &  1994  &&  b  &  AnB  & 0.8  & 1.3  \\
Sept &  15  &  1994  &&  c  &  BnC  & 3.2  & 4.0  \\
Sept &  28  &  1995  &&  d  &  AnB  & 4.2  & 1.2  \\
Oct  &   3  &  1992  &&  e  &   D   & 1.0  &29\phantom{.}\phantom{0}\\
\noalign{\smallskip}
\hline
\end{tabular}
\end{flushleft}
\end{table}

Except for the September 1995 experiment which used 3C48, absolute flux
calibration was set using 3C286 and the flux scale of \cite{baars:date} as
modified in the 15APR92 version of the Astronomical Image Processing System
(AIPS).  Phase calibrators were observed every few hours during each of the
experiments.

Data reduction consisted of making tapered $180{\arcsec} \times 180{\arcsec}$
CLEANed images and using only the first clean component to phase
self-calibrate the data (equivalent to using a point source model at the
location of the strongest radio source).  A second untapered map, centered at
the location of the peak on the first map, was made and CLEANed.  The rms
noise was measured in a region excluding all sources on the final map.  The
position and flux density of all sources whose signal-to-noise ratio exceeded
5 were recorded.

In Table 2 we present the 1861 RGB sources for which radio components were
detected.  We present only a sample page here; a full copy of the table is
available from the CDS via anonymous ftp to cdsarc.\linebreak[0]u-strasbg.fr
(130.79.128.5), via the WWW at http://cdsweb.\linebreak[0]u-strasbg.fr/
\linebreak[0]Abstract.html or at ftp://ftp.\linebreak[0]astro.\linebreak[0]psu.
\linebreak[0]edu\linebreak[0]/pub/\linebreak[0]edf\linebreak[0]/rgb\_tab2.html
and rgb\_tab3.html), or by contacting the authors.  The columns in Table 2 give
the source name, J2000 radio position, observation code (defined in Table 1),
signal-to-noise ratio, corrected 5\,GHz core VLA flux density (S$^{\rm
core}_{5}$), total 5\,GHz Green Bank flux density taken from the \cite{gnc2:g96}
catalog or from the reanalysis of the GB survey images (S$^{\rm tot}_{5}$), and
error of the total flux density if the source appeared in \cite{gnc2:g96}.  We
refer to individual sources using the catalog prefix ``RGB J'' (RASS-Green Bank
catalog, J2000 epoch positions) and append ``A'', ``B'', ``C'', etc.\ to denote
multiple radio sources found on a particular field.

\begin{table}
\caption[]{See Table at End of document}
\begin{flushleft}
\begin{tabular}{l}
\end{tabular}
\end{flushleft}
\end{table}

In Table 3 we show a typical page of similar information for the 436 sources
detected only at low resolution which have been excluded from further analysis.
Table 4 lists the 83 fields for which no source with a signal-to-noise ratio
greater than 5 was detected.  Many of these are faint sources cataloged by
\cite{neumann:date} but not in GB96 which used the stricter criterion for source
existence and are probably spurious.  The columns list the source name, J2000
Green Bank position, observation code, total GB 5\,GHz flux density, and error
in the total flux density if the source appeared in \cite{gnc2:g96}.  In
addition, two RGB sources (RGB J0425+179, RGB J1303+488) were not observed with
the VLA, but are part of the complete RGB sample.

\begin{table}
\caption[]{See Table at End of document}
\begin{flushleft}
\begin{tabular}{l}
\end{tabular}
\end{flushleft}
\end{table}

\begin{table}
\caption[]{See Table at End of document}
\begin{flushleft}
\begin{tabular}{l}
\end{tabular}
\end{flushleft}
\end{table}

The last column in the tables indicates the presence of a note which
indicates: (1) the source may be spurious or related to a diffuse Galactic
object (e.g. a supernova remnant); (2) the core radio flux density is from an
observation other than this paper; or (3) the source is more than 3$\sigma$
from its GB position (\S3.2).

\subsection{Flux Density Corrections} 
Instrumental effects degrade the measured flux density for sources far from
the field center.  While time average smearing is insignificant for our
observations, both bandwidth smearing (chromatic aberration) and primary beam
degradation are significant for many sources in the RGB catalog.  The
corrected flux density, ${\rm S}$, is given by:
\begin{equation}
S = \frac{S^{\prime}}{B\cdot P}
\end{equation}
where S$^{\prime}$ is the flux density from the final map.  The bandwidth
smearing correction term, ${\rm B}$, and the primary beam correction term, ${\rm
P}$, are given by (\cite{condon:fullnvss}):
\begin{equation}
{\rm B} = \left[1 + \frac{2~\ln 2}{3} \left(\frac{\Delta\nu~\rho}{\nu\Theta_0}
\right)^2 \right]^{-\frac{1}{2}} {\rm and}
\end{equation}
\begin{equation}
{\rm P} = \left( {\rm a}_0 + {\rm a}_1{\rm x} + {\rm a}_2{\rm x}^2 + {\rm a}_3
{\rm x}^3 + {\rm a}_4{\rm x}^4 \right)^{-1}.
\end{equation}
Here $\Delta\nu$ is the bandwidth (50\,MHz), $\nu$ is the observing frequency
(4.885\,GHz), $\rho$ is the angular distance from the field center, $\Theta_0$
is the restoring beam size (Table 1), ${\rm x}$\linebreak[0]=
\linebreak[0]$(\rho \nu)^2$ ($\rho$ in arcminutes), ${\rm a}_0$\linebreak[0]=
\linebreak[0]$1.003$, ${\rm a}_1$\linebreak[0]=
\linebreak[0]$1.086\times10^{-3}$, ${\rm a}_2$ \linebreak[0]=
\linebreak[0]$3.30\times10^{-6}$, ${\rm a}_3$\linebreak[0]=
\linebreak[0]$-3.609\times10^{-9}$ and ${\rm a}_4$\linebreak[0]=
\linebreak[0]$3.305\times 10^{-12}$.  

\subsection{Source Parameter Reliability}
While the formal uncertainties for our reported flux densities and positions
can be defined as a quadratic sum of the squares of several error terms (e.g.
\cite{cnc:full}; \cite{kollgaard:full}), we found these formal uncertainties
underestimated the true uncertainties in the reported source parameters.
The biggest sources of error in the RGB catalog are instead due to
instrumental and technical effects intrinsic to our snapshot mode.  In order
to assess the reliability in our measured flux densities and positions, we
observed 20 RGB sources at more than one epoch after the main survey was
completed and used the same data reduction procedure to obtain core flux
densities and positions.

These repeated observations show the reported positions for sources in Table 2
are accurate to $\le$$0.5{\arcsec}$ while those in Table 3 are accurate to
$\le$$8{\arcsec}$.  The core flux densities of the sources observed at
multiple epochs varied significantly, however, with the source intensity
varying by as much 80\% between epochs separated by as little as 10 days.
While some of this variability may be intrinsic to the sources, we
believe much of it is due to instrumental causes such as different VLA
resolution, the lack of phase calibrators near individual sources, very short
observation times, and consequently the small number of visibilities used to
image large fields.  Our tests show the reported flux densities of sources
fainter than $\sim$20~mJy are generally accurate to $\sim$50\% and the brighter
sources accurate to $\sim$20\%.

\section{Comparison of the Green Bank and VLA Source Properties}
\begin{figure}
\psfig{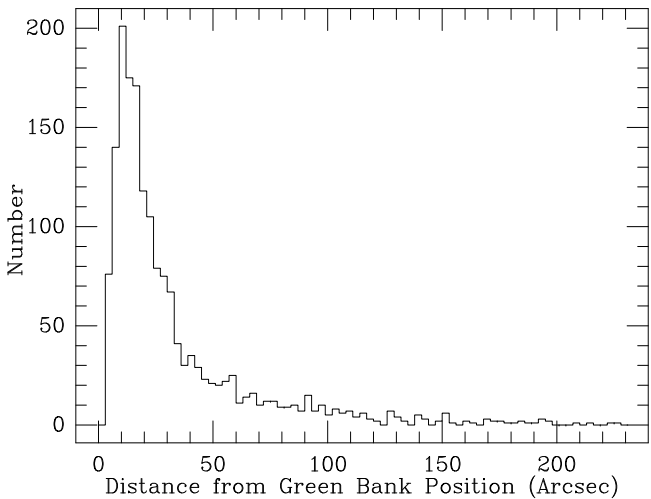}
\caption[]{Distribution of VLA - Green Bank Radio Positions}
\end{figure}
Figure 1 shows that 82\% of the sources are within $\sim$50$\arcsec$ of the
expected \cite{gnc2:g96} position and the distribution peaks at an offset of
10$\arcsec$-15$\arcsec$, consistent with the positional errors given in
\cite{neumann:date} and \cite{gnc:date}.  Because the positional uncertainty
of the Green Bank positions (which greatly dominate the uncertainties in the 
VLA positions) are flux dependent, we split the data into two subsets:  a
bright and faint sample with a division at 75~mJy.  We find that the
positional accuracies given in \cite{neumann:date} are then reliable for these
two samples.  For the analyses that follow, we therefore exclude all bright
sources offset from the GB positions by more than 45${\arcsec}$ and all faint
sources offset by more than 120${\arcsec}$.  These criteria exclude 74 objects
from the bright sample and 47 objects from the faint.  The number of spurious
coincidences which remain in the tables is quite small with only $\sim$31 of
the 757 remaining faint sources and $\sim$10 of the 813 remaining bright
sources expected to be spurious.  Flags are given in Table 2 to indicate the
sources we excluded.

\begin{figure}
\psfig{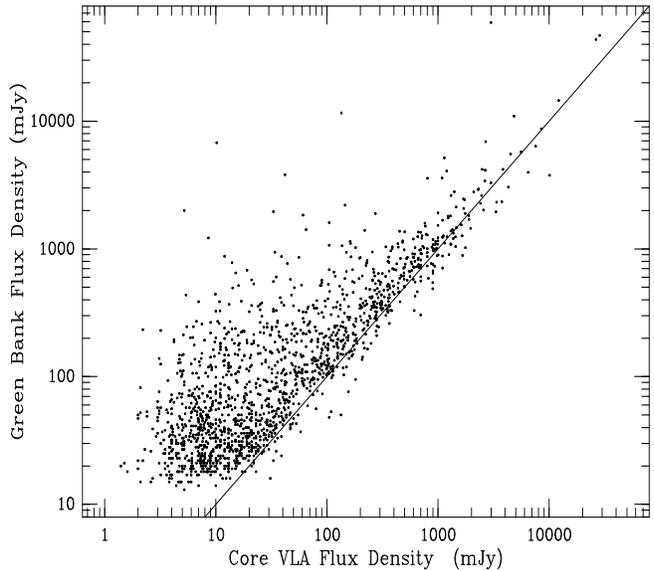}
\caption[]{Log-Log Diagram of the Core VLA Radio Flux Density (mJy) vs.\ Green
Bank Flux Density (mJy) at 5\,GHz}
\end{figure}

Figure 2 shows the flux-flux diagram comparing the core VLA flux density with
the Green Bank measurement.  For those $\sim$10\% of the fields for which more
than one source was detected, we have chosen as the ``core'' component that
source which is closest to the RASS position.  We also considered using the
brightest source on the field or the source closest to the GB position.  In
all the following analyses, the differences between the results obtained
using these three different criterion are well within the uncertainties in the
data.  Only when we completely excluded the fields with more than one detected
source did any of the results change significantly, increasing the median
core-to-lobe parameter (\S5) by $\sim$20\%.  While source variability and
uncertainties in both flux density measurements produces a few points in Figure
2 where ${\rm S}_{\rm 5}^{\rm VLA}$$>$${\rm S}_{\rm 5}^{\rm GB}$, it is clear
that nearly all of the sources in the GB catalog contain emission which is
significantly resolved at the arcsecond-scale.

\begin{figure}
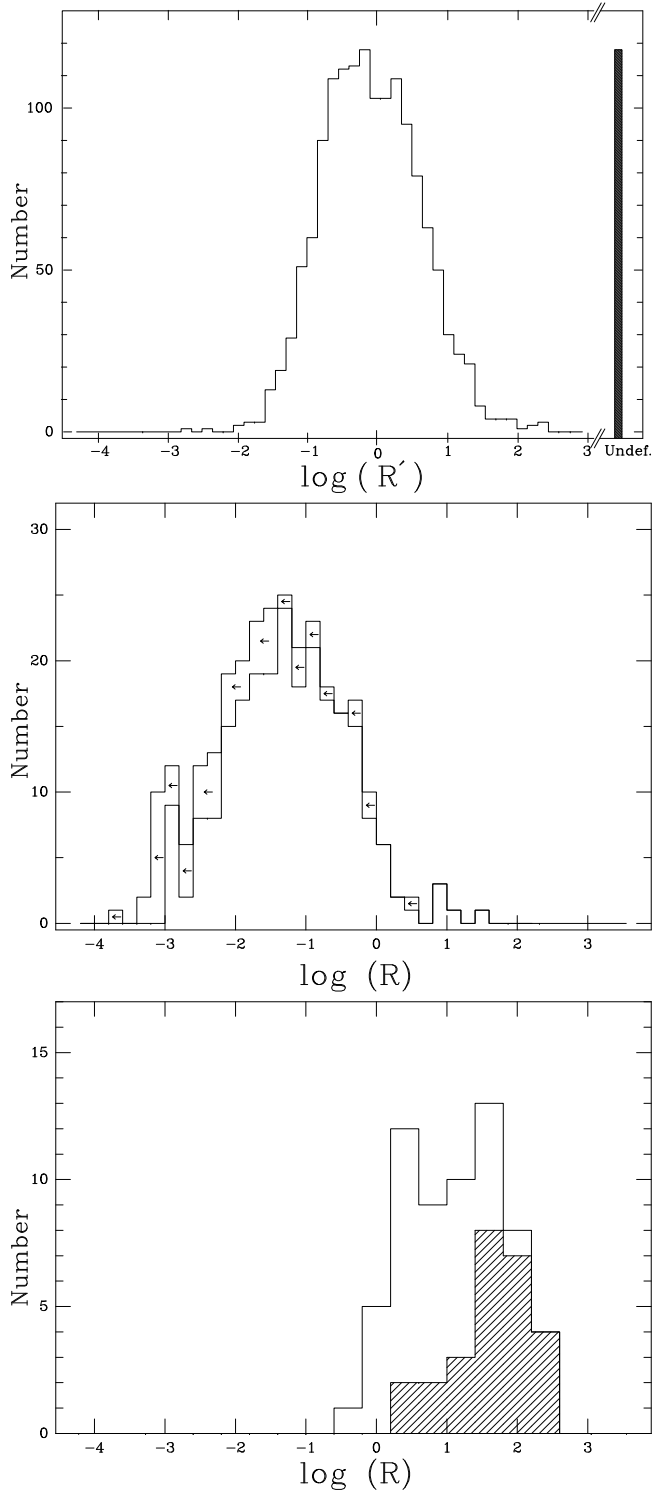

\psfig{file=c2l.epsi,height=6.5cm,width=8.5cm,angle=-90}
\vspace*{0.1cm}
\psfig{file=zirbel.epsi,height=6.5cm,width=8.5cm,angle=-90}
\vspace*{0.1cm}
\psfig{file=bll.epsi,height=6.5cm,width=8.5cm,angle=-90}
\caption[]{{\bf a.} Distribution of R$^{\prime}$, defined as the (VLA Core
Flux Density)/(Green Bank Flux Density - VLA Core Flux Density).  This
parameter is a rough measure of the radio core-to-lobe ratio.  The last bin
contains the 118 sources for which the Green Bank flux density is less than
the VLA core flux density.  {\bf b.} Distribution of the core-to-lobe ratio
for radio galaxies.  Bins containing undetected cores are denoted with a
left-facing arrow.  {\bf c.} Distribution of the core-to-lobe ratio for BL
Lacertae objects.  The hatched histogram distinguishes the radio--selected from
the X-ray--selected objects.}
\end{figure}

\section{Radio Beaming}
The radio core-to-lobe ratio, ${\rm R}$$=$${\rm S}^{\rm core}_{\rm 5}/{\rm
S}^{\rm lobe}_{\rm 5}$, is often used in studies of radio-loud AGN as a
relative measure of orientation (e.g. \cite{orrbrowne:full}).  In the absence
of high quality interferometric maps showing full details of the radio
structure, this ratio can be approximated as ${\rm R}^{\prime}$$=$${\rm S}^{\rm
VLA}_{\rm 5}/({\rm S}^{\rm GB}_{\rm 5} - {\rm S}^{\rm VLA}_{\rm 5})$.  The
distribution of ${\rm R}^{\prime}$ (Figure 3a) will differ in detail from that
of the true ${\rm R}$, both because the VLA and GB observations were not
simultaneous and because the low resolution GB measurement may contain some
emission from unrelated sources.  Nevertheless, we believe these effects should
not introduce any significant biases into the distribution of ${\rm
R}^{\prime}$.

For comparison we show in Figure 3b the distribution of core-to-lobe ratios
for a large sample of FRI and FRII radio galaxies compiled by \cite{zb:date}.  
(See \cite{zb:full} for a discussion of the assumptions used to derive ${\rm
R}$ and the upper limits on ${\rm R}$ for those sources without a measured
radio core.)  To illustrate the properties of an extremely core-dominated
population, we show in Figure 3c the core-to-lobe ratio for BL Lacertae
objects, with the radio-selected objects (RBLs) represented by the hatched
histogram.  The objects shown are the X-ray--selected BL Lacs (XBLs) from the
HEAO-1 Large Area Sky Survey (\cite{heao:full}) and the Einstein Extended Medium
Sensitivity Survey (\cite{morris:full}), and the radio--selected BL Lacs in the
1 Jansky sample (\cite{stickel:full}).  The radio flux densities used to derive
the core-to-lobe ratios were taken from Kollgaard et al.

Figure 3 shows the RGB sample is more core-dominated (40\% of the sources have
$\log$R$^{\prime}$$>$0) than the radio galaxy sample of Zirbel \& Baum (1995;
3\% with $\log$R$>$0), but is less core-dominated than the BL Lacertae objects
(82\% with $\log$R$>$0).  We used the Astronomy SURVival (ASURV) data analysis
software (Rev.\ 1.2; \cite{asurv:full}) to compute the Kaplan-Meier estimator
of the R distributions.  This properly takes into account the upper limits in
the radio galaxy sample (\cite{kmest:full}).  The median ${\rm R}$ of each
distribution is given in Table 5.  We find that both classes of BL Lac objects
are significantly more core-dominated than the RGB sample.  The median of
the radio galaxy sample, however, is 27 times less core-dominated than the RGB
sources.

The differences discussed above are clearly due to the type of object which
dominates each of the samples.  Although $>$70\% of the RGB catalog is optically
unidentified, most of the identified sources are quasars (B95).  A comparison of
the optically identified and unidentified sources shows that while the
identified sources generally exhibit higher radio and X-ray fluxes, other
properties (e.g.  their optical colors) are not statistically different (B96).
This suggests the unidentified sources are also primarily quasars.  The
differences in the distribution of ${\rm R}$ therefore indicate the RGB catalog
consists primarily of quasars whose radio emission is moderately beamed.

Within the framework of the unified scheme scenario which hypothesizes flat and
steep spectrum quasars are radio galaxies seen close to the line-of-sight (e.g.
\cite{barthel:full}), we use a simple beaming model and the core-to-lobe ratio
distributions to constrain the jet speed and orientation characteristic of
objects in the RGB sample.  The dependence of ${\rm R}$ on jet speed and
orientation are given by (e.g. \cite{unified:full}):
\begin{equation}
{R \equiv \frac{S^{\rm core}_{\rm r}}{S^{\rm lobe}_{\rm r}} = f \delta^p}
\end{equation}
where $f$ is the intrinsic core-to-lobe ratio, $p$ is the beaming index, and
$\delta$ is the Doppler factor:
\begin{equation}
{\delta = [\Gamma(1 - \beta \cos\theta)]^{-1}.}
\end{equation}
Here $\beta = {\rm v}/{\rm c}$, where ${\rm v}$ is the bulk velocity, $\Gamma
= (1 - \beta^2)^{-\frac{1}{2}}$, and $\theta$ is the angle to the
line-of-sight.  We assume $p = 2.7,$ applicable to a jet consisting of a
single sphere with a spectral index $\alpha$=0.3 (${\rm S}_{\nu} \propto
\nu^{\alpha}$; e.g. \cite{pnz:full}).  We make the further assumption that the
Zirbel \& Baum (1995) sample of FRI and FRII radio galaxies is characteristic of
the parent population of RGB sources, although we examine this hypothesis more
carefully at the end of this section.

\begin{table}
\caption[]{Median Core-to-lobe Ratios}
\begin{flushleft}
\begin{tabular}{ll}
\hline\noalign{\smallskip}
\multicolumn{1}{c}{Sample}&\multicolumn{1}{c}{${\rm R}$}\\
\noalign{\smallskip}
\hline\noalign{\smallskip}
Radio Galaxies&\phantom{2}0.019\\
RGB Sample&\phantom{2}0.52\\
XBLs&\phantom{2}1.8\\
RBLs&20.\\
\noalign{\smallskip}
\hline
\end{tabular}
\end{flushleft}
\end{table}

Kollgaard et al.\ (1996), analyzing the same population of radio galaxies and
BL Lacertae objects, found that the relative core enhancement of these
populations implied that $\Gamma$$>$4.5 and probably exceeded $\Gamma$=6.  We
therefore assume initially $\Gamma$=6 for all three populations and adopt
$\overline{\theta}$$=$$60^{\circ}$ for the radio galaxies.  (See Kollgaard et
al.) For a sample like the RGB catalog which consists largely of radio-loud
quasars (B95; B96; \cite{lm:full}), the assumption $\Gamma$=6 is a reasonable
lower limit to the jet speed (\cite{unified:full}).  Using these assumptions
and the median R values in Table 5, this implies that the average angle to the
line-of-sight for the RGB sample ($\overline{\theta}_{\rm RGB}$) is
approximately $30^{\circ}$, significantly larger than that obtained for the BL
Lac objects where $\overline{\theta}_{\rm XBL}$$\approx$$20^{\circ}$ and
$\overline{\theta}_{\rm RBL}$$\approx$$10^{\circ}$ (\cite{heao:full}).

\begin{figure}
\psfig{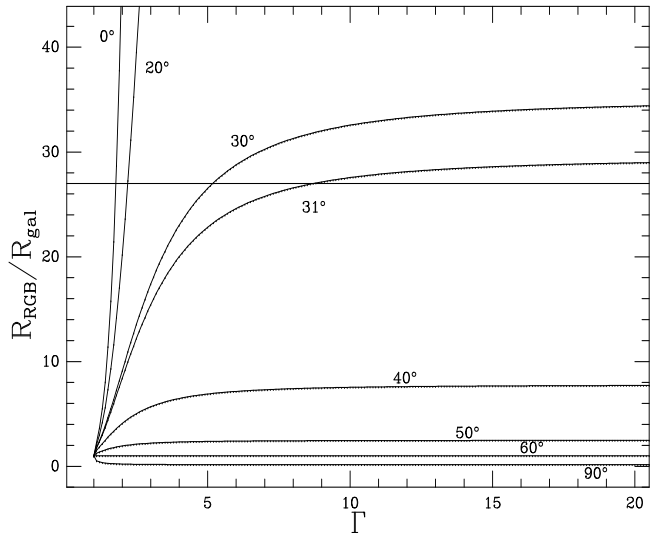}
\caption[]{The predicted line-of-sight orientation for the RGB sample,
$\overline{\theta}_{\rm RGB}$, derived from the relative median core-to-lobe
ratios of the RGB and radio galaxy samples.  $\overline{\theta}_{\rm
gal}=60^{\circ}$ is assumed and the horizontal line is the observed ratio of
$\frac{{\rm R}_{\rm RGB}}{{\rm R}_{\rm gal}}$.}
\end{figure}

The assumption that $\Gamma$ is a single value is most likely incorrect in
detail since a range of jet speeds probably characterizes any given population
of objects.  Assuming $\overline{\theta}_{\rm gal}$=$60^{\circ}$ and
$\Gamma_{\rm RGB}$$=$$\Gamma_{\rm gal}$, but allowing both Lorentz factors to
vary over the range 2$\ge$$\Gamma$$\ge$20, constrains the average angle to the
line-of-sight for the RGB sample to lie within a fairly small range,
20$^{\circ}$$<$$\overline{\theta}_{\rm RGB}$$<$32$^{\circ}$ (Figure 4).
If we further constrain $\Gamma$$\ge$5, which is a reasonable minimum based on
studies of the observed luminosity function of flat and steep spectrum quasars
(\cite{unified:full}), then $\overline{\theta}_{\rm RGB}$ is narrowly confined
to be about 31$^{\circ}$.

Finally, we consider the possibility that the population of FRI and FRII radio
galaxies used here is not characteristic of the parent population of objects
in the RGB catalog.  Assuming some form of a unified scheme is not
unreasonable, but it is possible that the RGB sample exhibits an average jet
speed substantially different than that characteristic of the Zirbel \& Baum
(1995) radio galaxy sample.  This could be the case if the RGB catalog is
biased toward objects with a larger $\Gamma$.  Using the results of
\cite{heao:date}, we fix $\Gamma_{\rm gal}$$=$6 and $\overline{\theta}_{\rm
gal}$$=$$60^{\circ}$.  As before, we constrain $\Gamma_{\rm RGB}$ to be larger
than 5.  The average angle to the line-of-sight for the RGB sample is then
20$^{\circ}$$\ge$$\overline{\theta}_{\rm RGB}$$\ge$35$^{\circ}$.  We also
consider the case where the intrinsic core-to-lobe ratio ({\it f} in Eqn. 4) of
the Zirbel \& Baum (1995) radio galaxies is different than that of the ``true''
parent population.   Since the RGB catalog is likely dominated by radio-- and
X-ray--loud quasars, if the FRII--quasar unified scheme is correct
(\cite{barthel:full}), the true parent population of RGB objects will have
extended radio powers approximately two to three orders of magnitude higher than
those objects in Zirbel \& Baum (1995).  Because the core-to-lobe ratio
decreases with increasing extended radio power (\cite{heao:full}), our ratio of
the core-to-lobe parameters would be too low by a factor of $\sim$4.  However,
the effect on the average angle to the line-of-sight is fairly modest,
decreasing it to $\overline{\theta}_{\rm RGB}$$\sim$25$^{\circ}$.  These
considerations indicate that $\overline{\theta}_{\rm RGB}$ is relatively
insensitive to assumptions about the detailed characteristics of the parent
population.

\section{X-ray Beaming?}
\begin{figure*}
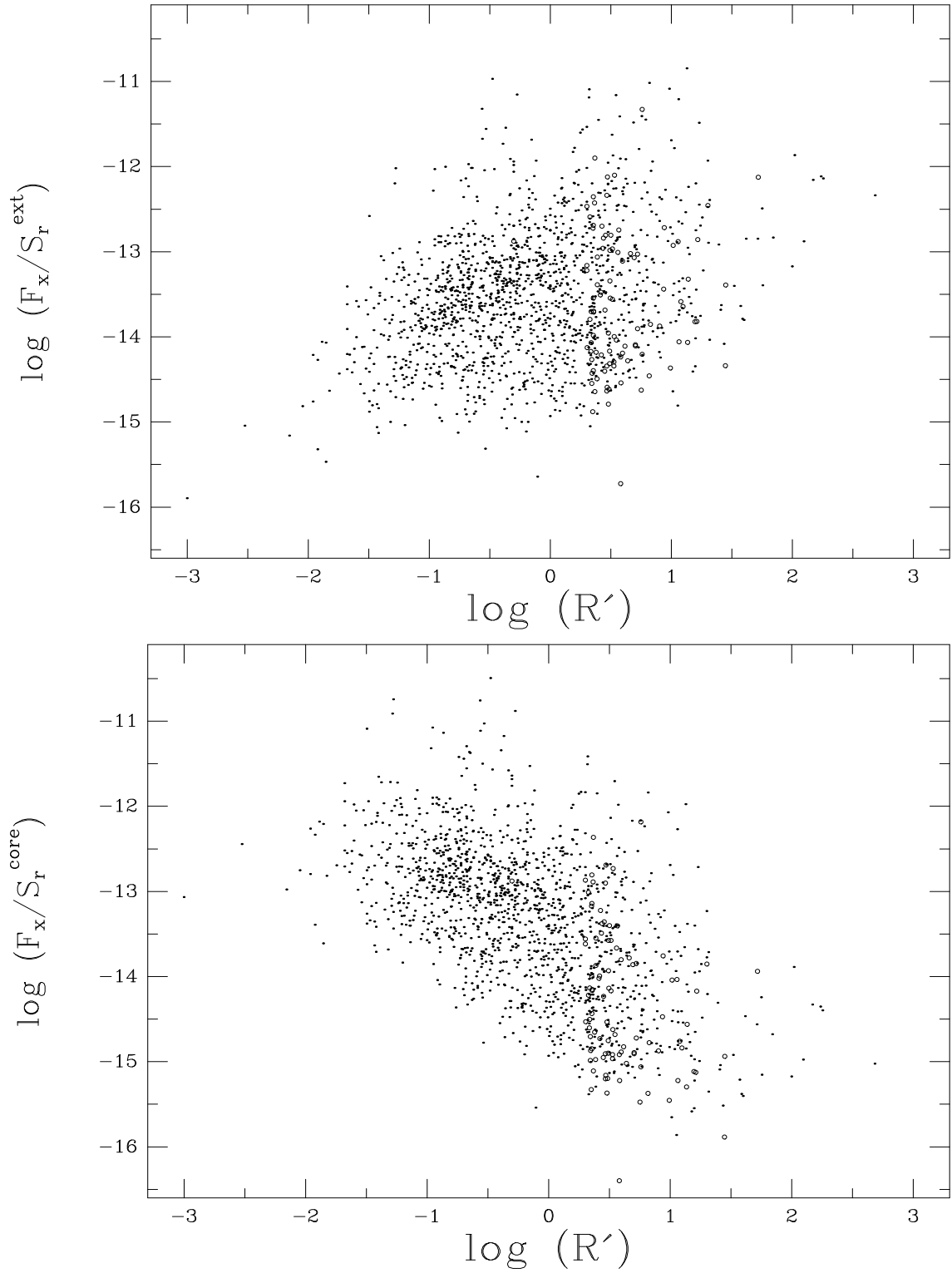

\psfig{file=xray2.epsi,height=10cm,width=15cm,angle=-90}
\vspace*{0.2cm}
\psfig{file=xray1.epsi,height=10cm,width=15cm,angle=-90}
\caption[]{{\bf a.} The ratio of X-ray to extended radio flux density as a
function of the radio core-to-lobe ratio, ${\rm R}^{\prime}$, for the RGB
sources.  For those 118 sources for which the VLA core flux density was
greater than the Green Bank total flux density, we derive upper limits as
described in the text.  These points are denoted by open circles in the
diagram.  {\bf b.} Similar to {\bf (a)} except the ordinate shows the ratio of
X-ray to core radio flux density.}
\end{figure*}

While the radio emission from radio-loud AGN consists of both beamed (core)
and unbeamed (extended) components, the origin of the X-ray emission is not as
clear.  Only recently has the deconvolution of thermal (host galaxy) X-ray
emission from nonthermal unresolved emission been possible (e.g.
\cite{wb:full}).  Although these studies are preliminary and the sample sizes
small, it is reasonable to assume the X-ray fluxes for the RGB sample likely
consist of a heterogeneous mix of these two components.  General trends should
nevertheless persist in the data revealing how much, if any, of the X-ray
emission is beamed.

Because redshifts are not available for most of the objects in the RGB sample,
flux ratios must be used to compensate for distance effects.  Specifically, we
compare the ratio of the X-ray flux (${\rm F}_{\rm x}$, in ${\rm
erg~s^{-1}~cm^{-2}}$) to the beamed radio core (${\rm S}_{\rm r}^{\rm core}$ in
mJy) and unbeamed extended (${\rm S}_{\rm r}^{\rm ext}$) components, with the
radio core-to-lobe ratio ${\rm R}^{\prime}$ (Figure 5).  For those sources
where our VLA core measurement exceeded the GB total flux density measurement,
we derive limits (in both the ordinate and abscissa) by assuming the maximum
uncertainty in the radio flux densities, namely 20\% error for sources
$>$20~mJy and 50\% error otherwise.  (See \S3.2.)  It is likely that the
uncertainty in the measurements and not source variability are to blame since
most of the sources which suffer from this effect are faint and therefore have
the greatest uncertainty in their measured flux densities.  Nevertheless,
because there are so many sources where ${\rm S}_{\rm r}^{\rm ext}$$\ge$${\rm
S}_{\rm r}^{\rm core}$, the statistical significance of the following analyses
remains unchanged whether the upper limits are halved or doubled, thus insuring
our results are insensitive to our particular method for computing the limits.

Figure 5a shows the ratio of the total X-ray flux to extended radio flux
density (${\rm F}_{\rm x}/{\rm S}_{\rm r}^{\rm ext}$) versus ${\rm
R^{\prime}}$ while in Figure 5b we show the ratio of the total X-ray flux to
core radio flux density (${\rm F}_{\rm x}/{\rm S}_{\rm r}^{\rm core}$) versus
${\rm R}^{\prime}$.  Two trends are seen in Figure 5: ${\rm F}_{\rm x}/{\rm
S}^{\rm ext}_{\rm r}$ increases with increasing ${\rm R}^{\prime}$, and ${\rm
F}_{\rm x}/{\rm S}^{\rm core}_{\rm r}$ decreases with increasing ${\rm
R}^{\prime}$.  Both these trends are statistically significant at the
$>$99.99\% level.  Two possible biases could affect the correlations.  First,
if the 83 sources for which no arcsecond-scale source was detected (Table 4)
are real and not spurious detections in the 3$\sigma$ Green Bank catalog, they
must be lobe-dominated and would appear on the left-hand side of Figure 5a.  If
these sources were also systematically X-ray brighter so that they had high
${\rm F}_{\rm x}/{\rm S}^{\rm ext}_{\rm r}$ ratios then they could populate the
upper left portion of Figure 5a; however, the ROSAT fluxes for these sources
span the same range as the detected sources, indicating this potential bias is
not present.  Second, the different flux limits of the original GB catalog
(${\rm S}^{\rm tot}_{\rm r}$$\ge$15~mJy) and the deeper VLA core measurements
($\sim$1~mJy) imply core-dominated sources with 1$\le$${\rm S}^{\rm tot}_{\rm
r}$$\le$15~mJy are missing from the RGB catalog.  These ``missing'' sources
could destroy the correlation in Figure 5b only if their core-to-lobe ratios
exceeded $\sim$10.0 and ${\rm F}_{\rm x}$ exceeded
$\sim$$10^{-13}$~erg~s$^{-1}$~cm$^{-2}$.  This is not the case, however, since
our VLA flux limit is only one order of magnitude deeper than the GB flux limit
thereby constraining the core-to-lobe ratios of the missing sources to be
-1.0$<$$\log {\rm R}^{\prime}$$<$1.0.  We therefore conclude that the two
trends in Figure 5 are real.

To understand these relationships, we characterize the X-ray emission by an
X-ray ``core-to-extended'' ratio, ${\rm R}_{X}$, defined by:
\begin{equation}
{{\rm R}_{X} \equiv \frac{{\rm F}_{\rm x}^{\rm core}}{{\rm F}_{\rm x}^{\rm
ext}}},
\end{equation}
If the X-ray core beaming is simply related to the radio core beaming, we can
write ${\rm R}_{\rm X}$ =  $k {\rm R}^{\prime}$ where {\it k} is a constant.
The quantities plotted in Figure 5 are then:
\begin{equation}
\frac{{\rm F}_{\rm x}}{{\rm S}^{\rm core}_{\rm r}} = \frac{{\rm F}^{\rm ext}
_{\rm x}}{{\rm S}^{\rm ext}_{\rm r}} \frac{1 + k {\rm R}^{\prime}}{{\rm
R}^{\prime}}
\end{equation}
and
\begin{equation}
\frac{{\rm F}_{\rm x}}{{\rm S}^{\rm ext}_{\rm r}} = \frac{{\rm F}^{\rm ext}
_{\rm x}}{{\rm S}^{\rm ext}_{\rm r}} (1 + k {\rm R}^{\prime}).
\end{equation}
First we consider the case where the X-ray emission is isotropic so that
$F_{\rm x}$$=$$F_{\rm x}^{\rm ext}$ and $k$=0.  Then as the angle to the
line-of-sight decreases, both ${\rm R}^{\prime}$ and ${\rm S}_{\rm r}^{\rm
core}$ increase but ${\rm F}_{\rm x}$ remains constant.  The ratio of ${\rm
F}_{\rm x}/{\rm S}^{\rm core}_{\rm r}$ would therefore be anticorrelated with
${\rm R}^{\prime}$ as seen in Figure 5b.  However, the ratio of ${\rm F}_{\rm
x}/{\rm S}^{\rm ext}_{\rm r}$ would be uncorrelated with ${\rm R}^{\prime}$
since neither parameter would vary with orientation.  The positive correlation
in Figure 5a, therefore rules out the possibility that X-ray emission for
sources in the RGB catalog is entirely isotropic.

If we now consider the other extreme where the X-ray emission consists of a much
higher fraction of beamed radiation than the radio emission ($k$$>>$1).  ${\rm
F}_{\rm x}/{\rm S}^{\rm ext}_{\rm r}$ should then be correlated with ${\rm
R}^{\prime}$, as observed, but the ratio of ${\rm F}_{\rm x}/{\rm S}^{\rm
core}_{\rm r}$ would become uncorrelated with ${\rm R}^{\prime}$ at even modest
values of ${\rm R}^{\prime}$, which is clearly not seen (Figure 5b).  Therefore,
if the X-ray emission is beamed, it is not characterized by a high {\it
k}-value.  As an alternative to the models presented in Equations 6-8, we consider
the case where $\Gamma_{\rm x}$$\ne$$\Gamma_{\rm r}$.  Such a scenario has been
proposed in terms of an accelerating jet model for BL Lac objects (e.g.
\cite{jet:full}) where $\Gamma_{\rm x}$$\le$$\Gamma_{\rm r}$.  We find that in
order to produce the relations seen in Figure 5 which are valid over three orders
of magnitude in the X-ray to radio flux ratios and five orders of magnitude in
${\rm R}^{\prime}$, $\Gamma_{\rm x}$$\ge$4 and $\Gamma_{\rm r}$$\ge$6, which is
consistent with bulk velocities inferred through other means (e.g.
\cite{unified:full}).

Figures 5a and 5b therefore indicate the X-ray emission of the RGB sample is
neither entirely isotropic (${\rm R}_{\rm X}$=0) nor characterized by a high
{\it k}-value.  However, the scatter in the diagrams is large enough to
prevent an accurate measurement of the fraction of beamed X-rays or even to
distinguish beamed X-ray emission from unbeamed but anisotropic emission.  The
latter could arise, for example, from a population of objects with an
obscuring torus with varying column density which blocks more soft X-rays as
the torus becomes more edge-on to the line-of-sight.

The large scatter in the diagrams is primarily
due to the heterogeneity of the RGB sample, which includes radio galaxies,
quasars and BL Lacs.  Examination of a single class of AGN, such as RGB BL
Lacs, can yield more quantitative results (\cite{lm:full}).

\section{Conclusions}
We present subarcsecond radio positions and core radio flux densities for all
2,127 sources appearing in both the Green Bank 5\,GHz and in the ROSAT All-Sky
Survey catalogs.  The accuracy of the positions is sufficient to give unique
optical identifications for the X-ray-- and radio--emitting sources (B96).

This RGB sample is comprised principally of radio-loud AGN.  It is complete
and unbiased with well-defined selection criteria:  X-ray flux above the RASS
sensitivity limit (which depends only on ecliptic latitude), arcminute-scale
radio flux density above $\simeq$15~mJy at 5\,GHz, and declination between
0$^{\circ}$ and 75$^{\circ}$.

The radio emission of the RGB sample is found to be more core-dominated than
ordinary radio galaxies but less than strongly beamed BL Lac objects, which
suggests it consists primarily of moderately beamed AGN.  Using simple beaming
models, the typical RGB object is shown to be dominated by a jet oriented at an
intermediate angle to the line-of-sight ($\overline{\theta}_{\rm
RGB}$$\sim$25$^{\circ}$-35$^{\circ}$).  The X-ray beaming properties are not
tightly constrained, but exclude the extremes of purely isotropic emission and
of X-ray emission which consists of a significantly higher fraction of beamed
flux than is characteristic of the radio emission.

\begin{acknowledgements} We wish to thank C.\ Palma for help with the data
reduction at Penn State.  This work was partially supported by NASA under Grant
NAGW-2120.  We have made use of the NASA/IPAC Extragalactic Database, operated
by the Jet Propulsion Laboratory, California Institute of Technology, under
contract with NASA.
\end{acknowledgements}

\vspace*{1.0in}
\noindent{\bf Notes to Table 2}\\
\begin{scriptsize}
\noindent 1545+646A: More than 3$\sigma$ from GB position\\
1545+646B: More than 3$\sigma$ from GB position\\
1547+208: Core radio flux from Miller et al.\, 1993\\
1549+026: Core radio flux from Murphy et al.\, 1993\\
1550+113: Core radio flux from Hintzen et al.\, 1983\\
1604+012: Core radio flux from Baum \& Heckman 1989\\
1608+104: Core radio flux from Murphy et al.\, 1993\\
1609+179: Core radio flux from Hutchings et al.\, 1988\\
1617+350: More than 3$\sigma$ from GB position\\
1620+176: Core radio flux from Hintzen et al.\, 1983\\
1625+268: Core radio flux from Feigelson et al.\, 1984\\
\end{scriptsize}

\noindent{\bf Notes to Table 3}\\
\begin{scriptsize}
\noindent 0242+083: More than 3$\sigma$ from GB position\\
\end{scriptsize}

\noindent{\bf Notes to Table 4}\\
\begin{scriptsize}
\noindent 0140+087: Possibly Spurious Source; See Condon et al.\ 1994\\
0205+648: Extended Source\\
0247+187: Extended Source\\
0317+415: Confused Field - Possibly Spurious Source\\
0528+344: Extended Source\\
0533+210: Confused Field - Possibly Spurious Source\\
0535+222: Confused Field - Possibly Spurious Source\\
0641+080: Extended Source\\
1232+123: Confused Field - Possibly Spurious Source\\
1728+086: Confused Field - Possibly Spurious Source\\
1852+006: Extended Source (known SNR)\\
1859+071: Confused Field or Extended Source\\
2015+386: Confused Field - Possibly Spurious Source\\
2212+589: Extended Source\\
2302+587: Extended Source\\
2323+584: Confused Field - Possibly Spurious Source\\
\end{scriptsize}


\begin{thebibliography}{}
\bibitem[Baars et al. (1977)]{baars:date}
\bibitem[Baars et al. 1977]{baars:full}
Baars, J. W. M., Gnezel, R., Pauliny-Toth, I. I. K. \& Witzel, A. 1977,
A\&A, 61, 99
\vspace{-11pt}
\bibitem[Bade et al. (1995)]{bade:date}
\bibitem[Bade et al. 1995]{bade:full}
Bade, N., Fink, H. H., Engels, D., Voges, W., Hagen, H. J., Wisotzki, L. \&
Reimers, D., 1995, A\&AS, 110, 469\newline
\vspace{-11pt}
\bibitem[Barthel 1989]{barthel:full}
Barthel, P. D., 1989, ApJ, 336, 606\newline
\vspace{-11pt}
\bibitem[Baum \& Heckman 1988]{BH:full}
Baum, S. A. \& Heckman, T., 1989, ApJ, 336, 681
\vspace{-11pt}
\bibitem[Becker et al. 1991]{becker:full}
\bibitem[Becker et al. (1991)]{becker:date}
Becker, R. H., White, R. L. \& Edwards, A. L., 1991, ApJS, 75, 1
\vspace{-11pt}
\bibitem[Brinkmann et al. (1994)]{molonglo:date}
\bibitem[Brinkmann et al. 1994]{molonglo:full}
Brinkmann, W., Siebert, J. \& Boller, Th., 1994, A\&A, 281, 355
\vspace{-11pt}
\bibitem[Brinkmann et al. (1995, B95)]{Brinkmann:dateids_paperI}
\bibitem[Brinkmann et al. 1995]{Brinkmann:fullids}
Brinkmann, W., Siebert, J., Reich, W., F\"urst, E., Reich, P., Voges, W.,
Tr\"umper, J. \& Wielebinski, R., 1995, A\&AS, 109, 147 (B95)
\vspace{-11pt}
\bibitem[Brinkmann et al. (B96)]{siebert:datepaperIII}
\bibitem[Brinkmann et al. (1996; B96)]{siebert:firstpaperIII}
Brinkmann, W., Siebert, J., Feigelson, E. D., Kollgaard, R. I.,
Laurent-Muehleisen, S. A. \& McMahon, R., 1996, (in preparation; B96)

\bibitem[Browne \& Murphy 1987]{BM:full}
Browne, I. W. A. \& Murphy, V., 1987, MNRAS, 226, 601

\bibitem[Burns et al. 1984]{BBD:full}
Burns, J. O., Basart, J. P., De Young, D. S. \& Ghiglia, D. C., 1984, ApJ,
283, 515
\vspace{-11pt}
\bibitem[Condon et al. (1982)]{cnc:date}
\bibitem[Condon et al. 1982]{cnc:full}
Condon, J. J., Condon, M. A. \& Hazard, C. 1982, AJ, 87, 739
\vspace{-11pt}
\bibitem[Condon et al. (1996)]{condon:datenvss}
\bibitem[Condon et al. 1996]{condon:fullnvss}
Condon, J. J., Cotton, W. D., Greisen, E. W., Yin, Q. F., Perley, R. A. \&
Broderick, J. J., 1996, AJ, in preparation
\vspace{-11pt}
\bibitem[Elvis et al. 1992]{elvis:full}
\bibitem[Elvis et al. (1992)]{elvis:date}
Elvis, M., Plummer, D., Schachter, J. \& Fabbiano, G., 1992, ApJS, 80, 257\newline
\vspace{-11pt}
\bibitem[Feigelson et al. 1984]{FI:full}
Feigelson, E. D., Isobe, T. \& Kembhavi, A., 1984, AJ, 89, 1464\newline
\vspace{-11pt}
\bibitem[Feigelson \& Nelson 1985]{kmest:full}
Feigelson, E. D. \& Nelson, P. I., 1985, ApJ, 293, 192
\bibitem[Ghisellini \& Maraschi 1989]{jet:full}
Ghisellini, G. \& Maraschi, L., 1989, ApJ, 340, 181
\bibitem[Ghisellini et al.\, 1993]{GHI:full}
Ghisellini, G., Padovani, P., Celotti, A., Maraschi, L., 1993, ApJ, 407, 65
\vspace{-11pt}
\bibitem[Gioia et al. 1990]{gioia:full}
\bibitem[Gioia et al. (1990)]{gioia:date}
Gioia, I., Maccacaro, T., Schild, R., Wolter, A., Stocke, J., Morris, S. \&
Henry, J. P., 1990, ApJS, 72 567\newline
\vspace{-11pt}
\bibitem[Gower \& Hutchings 1984]{GH:full}
Gower, A. C. \& Hutchings, A. C., 1984, AJ, 89, 1658
\vspace{-11pt}
\bibitem[Gregory \& Condon (1991)]{gnc:date}
\bibitem[Gregory \& Condon 1991]{gnc:full}
Gregory, P. C. \& Condon, J. J. 1991, ApJS, 75, 1011
\vspace{-11pt}
\bibitem[Gregory et al.\ 1996]{gnc2:full}
\bibitem[GB96]{gnc2:g96}
Gregory, P. C., Scott, W. K., Douglas, K. \& Condon, J. J. 1996, ApJS,
submitted (GB96)\newline
\vspace{-11pt}
\bibitem[Hintzen et al. 1983]{HIN:full}
Hintzen, P., Ulvestad, J. \& Owen, F., 1983, AJ, 88, 709\newline
\vspace{-11pt}
\bibitem[Hutchings et al.\, 1988]{HPG:full}
Hutchings, J. B., Price, R. \& Gower, A. C., 1988, ApJ, 329 122\newline
\vspace{-11pt}
\bibitem[Kellerman et al.\, 1989]{KSS:full}
Kellerman, K. I., Sramek, R. A., Schmidt, M., Shaffer, D. B. \& Green, R. F., 1989,
AJ, 98, 1195\newline
\vspace{-11pt}
\bibitem[Kollgaard et al.\, 1992]{KWRG:full}
Kollgaard, R. I., Wardle, J. F. C., Roberts, D. H. \& Gabuzda, D. C., 1992, AJ,
104, 1687
\vspace{-11pt}
\bibitem[Kollgaard et al. 1994]{kollgaard:full}
\bibitem[Kollgaard et al. (1994)]{kollgaard:date}
Kollgaard, R. I., Brinkmann, W., Chester, M. M., Feigelson, E. D., Hertz, P.,
Reich, P. \& Wielebinski, R., 1994, ApJS, 93, 145
\vspace{-11pt}
\bibitem[Kollgaard et al. 1996]{heao:full}
\bibitem[Kollgaard et al. (1996)]{heao:date}
Kollgaard, R. I., Palma, C., Laurent-Muehleisen, S. A., Feigelson, E. D.,
1996, ApJ, To Appear
\vspace{-11pt}
\bibitem[Laurent-Muehleisen et al.\,, in preparation]{lm:full}
\bibitem[Laurent-Muehleisen et al. (in preparation)]{lm:date}
Laurent-Muehleisen, S. A., Kollgaard, R. I., Ciardullo, R. B., Feigelson, E.
D., 1996, in preparation\newline
\vspace{-11pt}
\bibitem[Lawson et al.\,, 1992]{LTW:full}
Lawson, A. J., Turner, M. J. L., Williams, O. R., Stewart, G. C., Saxton, R. D.,
1992, MNRAS, 259, 743
\vspace{-11pt}
\bibitem[LaValley et al.\,, 1992]{asurv:full}
\bibitem[LaValley et al.\,, (1992)]{asurv:date}
LaValley, M., Isobe, T. \& Feigelson, E. D., 1992, BAAS, 24, 839\newline
\vspace{-11pt}
\bibitem[Linfield \& Perley 1984]{LP:full}
Linfield, R. \& Perley, R., 1984, ApJ, 279, 60\newline
\vspace{-11pt}
\bibitem[Miller et al.\,, 1993]{MRS:full}
Miller, P., Rawlings, S. \& Sanders, R., 1993, MNRAS, 263, 425
\vspace{-11pt}
\bibitem[Morris et al. 1991]{morris:full}
\bibitem[Morris et al. (1991)]{morris:date}
Morris, S. L., Stocke, J. T., Gioia, I. M., Schild, R. E., Wolter, A., 
Maccacaro, T. \& Della Ceca, R., 1991, ApJ, 380 49\newline
\vspace{-11pt}
\bibitem[Murphy et al.\, 1993]{MBP:full}
Murphy, D. W., Browne, I. W. A. \& Perley, R. A., 1993, MNRAS, 264, 298
\vspace{-11pt}
\bibitem[Neumann et al. 1994]{neumann:full}
\bibitem[Neumann et al. (1994)]{neumann:date}
Neumann, M, Reich, W., F\"urst, E., Brinkmann, W., Reich, P., Siebert, J.,
Wielebinski, R. \& Tr\"umper, J., 1994, A\&AS, 106, 303\newline
\vspace{-11pt}
\bibitem[O'Dea \& Owen, 1985]{ODO:full}
O'Dea, C. P. \& Owen, F. N., 1985, AJ, 90, 927\newline
\vspace{-11pt}
\bibitem[Orr \& Browne, 1982]{orrbrowne:full}
Orr, M. J. \& Browne, I. W. A., 1982, MNRAS, 200, 1067
\vspace{-11pt}
\bibitem[Pearson \& Zensus 1987]{pnz:full}
\bibitem[Pearson \& Zensus (1987)]{pnz:date}
Pearson, T. J. \& Zensus, J. A., 1987, in Superluminal Radio Sources, Ed. J.
A. Zensus \& T. J. Pearson (Cambridge University Press, Cambridge), 1\newline
\vspace{-11pt}
\bibitem[Perley 1982]{perley:full}
Perley, R. A., 1982, AJ, 87, 859\newline
\vspace{-11pt}
\bibitem[Punsly 1995]{PUN:full}
Punsly, B., 1995, AJ, 109, 1555
\vspace{-11pt}
\bibitem[Stickel et al. 1991]{stickel:full}
\bibitem[Stickel et al. (1991)]{stickel:date}
Stickel, M., Padovani, P., Urry, D. M., Fried, J. W. \& K\"uhr, H., 1991, ApJ,
374, 431
\vspace{-11pt}
\bibitem[Stocke et al. 1991]{stocke:full}
\bibitem[Stocke et al. (1991)]{stocke:date}
Stocke, J. T., Morris, S. L., Gioia, I. M., Maccacaro, T., Schild, R., Wolter,
A., Fleming, T. A. \& Henry, J. P., 1991, ApJS, 76, 813\newline
\vspace{-11pt}
\bibitem[Ulvestad \& Wilson 1984]{UW:full}
Ulvestad, J. S. \& Wilson, A. S., 1984, ApJ, 278, 544
\vspace{-11pt}
\bibitem[Urry \& Padovani 1995]{unified:full}
\bibitem[Urry \& Padovani (1995)]{unified:date}
Urry, C. M. \& Padovani, P., 1995, PASP, 107, 803
\vspace{-11pt}
\bibitem[Voges et al. 1992]{voges92:full}
\bibitem[Voges et al. (1992)]{voges92:date}
Voges, W., Gruber, R., Paul, J., Bickert, K., Bohnet, A., Bursik, J., Dennerl,
K., Englhauser, J., Hartner, G., Jennert, W., K\"ohler, H. \& Rosso, C., 1992,
The ROSAT Standard Analysis Software System, ESA ISY-3, p. 223
\vspace{-11pt}
\bibitem[Voges 1993]{voges:full}
\bibitem[Voges (1993)]{voges:date}
Voges, W., 1993, Adv. Space Res., Vol. 13, No. 12, 391
\vspace{-11pt}
\bibitem[Wood et al. 1984]{wood:full}
\bibitem[Wood et al. (1984)]{wood:date}
Wood, K. S., et al., 1984, ApJS, 56 507
\vspace{-11pt}
\bibitem[Worrall \& Birkinshaw 1994]{wb:full}
\bibitem[Worrall \& Birkinshaw (1994)]{wb:date}
Worrall, D. M. \& Birkinshaw, M., 1994, ApJ, 427, 134
\vspace{-11pt}
\bibitem[Wrobel \& Heeschen 1984]{WH:full}
\bibitem[Wrobel \& Heeschen (1984)]{WH:date}
Wrobel, J. M. \& Heeschen, D. S., 1984, ApJ, 287, 41
\vspace{-11pt}
\bibitem[Zirbel \& Baum 1995]{zb:full}
\bibitem[Zirbel \& Baum (1995)]{zb:date}
Zirbel, E. L. \& Baum, S. A., 1995, ApJ, 448, 521
\end{thebibliography}
\end{document}